\documentstyle[12pt,a4]{article}

\newcommand{\be}{\begin{equation}}
\newcommand{\ee}{\end{equation}}

\date{\ }
\begin{document}
\def\R{{\hbox{{\rm I}\kern-. 2em\hbox{\rm R}}}}
\def\cP{{\cal P}}
\def\cR{{\cal R}}
\def\cW{{\cal W}}
\def\cte{{\mbox{\rm c}^{\mbox{\rm te}}\; }}
\begin{titlepage}
\begin{flushright} UMH-MG-4
\end{flushright}
\vskip 1. cm
\begin{center} {\LARGE\bf   Quantum Minimal Length and Transplanckian Photons  }\\
\vskip 1. cm M. Lubo\footnote{E-mail lubo@sun1. umh. ac. 
be}$\mbox{}^{\mbox{\footnotesize }}$, 
\\ {\em M\'ecanique et Gravitation, Universit\'e de Mons-Hainaut}, 
\\ {\em 6, avenue du Champ de Mars, B-7000 Mons, Belgium}
 \\
\vskip 0.5 cm
\vskip 1 cm
\end{center}
\begin{abstract}
 
In this paper,we first give arguments supporting the idea that a B.T.Z black hole can face a transplanckian problem even when its mass is small. K.M.M  quantum theory is  applied to the Hawking evaporation of
Schwarzchild and B.T.Z black holes.Working in the physically safe quasi position representation,
we argue that the oscillating term present in a previous analysis is removed so that actually
one doesn't need an average procedure.We  expand the s  wave  function
 as the exponential of a series in the minimal length of the new quantum theory.This reduces an infinite order differential 
equation to a numerable set of finite order ones.We
obtain the striking result that the infinity of arbitrary constants induced by the order of
the wave equation has no physical meaning due to normalisation.We finally construct
gaussian wave paquets and study  their trajectories .We  suggest a quantitative description
of the non locality zone and its dependance on the K.M.M energy scale. Potential incidences on unitarity are briefly evoked .
\end{abstract}
\vfill 
\end{titlepage}

\newpage 
     
\section{Introduction}
    Hawking effect \cite{Haw}is of crucial importance for black hole physics.Classically those objects are stable and don't
emitt any signal:they can be detected only through their strong gravitational fields.However,at the quantum level
the situation is radically different.Due to inequivalent vacuua interplay,black hole emitt energy.For example,the emission
of massless charged scalars in the Scharzchild geometry is thermal,at a temperatire proportional to the inverse of the 
 black hole mass.Due to this phenomenon,a black hole is no more eternal after its birth:its life time goes like the cube 
of its mass.
   This emission nicely feets the entropy and the temperature predicted by
Bekenstein \cite{Bek}. In spite of its inner beauty imbricating classical general relativity and quantum field theory,the  Hawking emission is
plagued by two major problems.The first one is unitarity:it is not obvious that one can succesfully handle a situation in which the source of gravity(the black hole) and the partner of the emitted particle finally merge into the singularity.The second problem is the transplanckian one :to be detected in the asymptotic region at the predicted momentum,the emitted photon should cross the star surface with a very big momentum.At this scale it is quite evident that the approximation by a
noninteracting scalar field is not reliable.In the begining of its journey,the Hawking photon also hugs the horizon at
incredibly small distances.
   In this paper we adress the second question.A truncation scheme modifying the dispersion relation was initiated by \cite{Ja},\cite{Un1} ,\cite{Un2} and also used by \cite{Br1} to cure the desease.A  recent tentative \cite{Br2} is a priori more fundamental               
since the change in the dispersion relation is not put by hand but appears as a consequence of  a "new" quantum theory. K.M.M approach \cite{Kem1},\cite{Kem2},\cite{Kem3} changes the rules of quantum mechanics in a very simple way and obtains a theory in which the uncertainty in position admitts a lower bound.One can then anticipate that this limit to localisability will forbid the Hawking particle to remain too close to the horizon. 
   A first attempt in this way raised a conceptual problem .Working in the momentum representation of the K.M.M  commutation relations,the s wave mode was calculated.Going to position representation by a Fourrier like tranform,it was realised that
the result did not satisfy the wave equation !It was then argued that this equation had to be verified only in average but the average procedure is still missing.
  In this paper we apply K.M.M formalism more carefully and restrict ourselves to physically meaningfull representations
.
The most sallient point is that in the "new" quantum quantum mechanics,states of definite positions do not exist.Their energy is infinite,they are not physical . Consequently,position representation doesn't exist and  the equation we tried to interpret is physically irrelevant.The
"unproblematic" quasi position representation is the best one for our purposes.When the minimal length uncertainty goes to
zero,it reduces to the well known position representation.We use it in the $3+1 $Schwarzchild solution and the $2+1$ B.T.Z
geometry \cite{BTZ1},\cite{BTZ2}.The behaviour of the modes near the horizon gratify our wishes to the full .  The paper is organised as follows.In the first section we focus on B.T.Z black holes  and the quantitative formulation
of its planckian problem.In the second  section we review the fundamentals of K.M.M theory and insist on the allowed representations.We then  criticise \cite{Br2} and  derive more physically motivated  wave equations. We handle 
the quasi position wave function in a perturbative way and finally construct the wave paquets.        
\section{ Minimally Coupled Scalar in BTZ }
    
\subsection{Introduction}
The introduction of a cosmological constant in Einstein equations
\[
G_{\mu\nu} = 8\pi G T_{\mu\nu} - \Lambda g_{\mu\nu}
\]
in a $ 2 + 1 $ space -time  generates, for $\Lambda >0$, the BTZ \cite{BTZ1},\cite{BTZ2} metric which,for a vanishing angular momentum,reads
\be
ds^2 =  (\Lambda r^2-M) dt^2 -{dr^2\over{\Lambda r^2-M}} - r^2d\theta^2
\ee
and is a solution for the field around a point source with an event
horizon. 
\par This black hole can be obtained by the collapse of a pressureless dust
surrounded by a vacuum region \cite{Mann}. The horizon and the surface of the collapsing star coincide at finite comoving
time   but at infinite exterior time .
\par To better apprehend the BTZ geometry, let us study the classical motion
of a particle in its background. As $t$ and $\theta$ are cyclic coordinates, one obtains two constants of 
motion : the energy $\tilde{E}$ and the angular momentum $\tilde{L}$. The on shell mass condition
can be rewritten as
\be
({dr\over{d\tau}})^2 = \tilde E^2 - V^2_{eff}(r)
\ee
 the potential governing the radial motion
being 
\be
V^2_{eff} (r) = \Lambda \tilde L^2-M + \Lambda r^2-{M\tilde L^2\over {r^2}}
\ee
It is a strictly increasing function. As $V^2_{eff}(\infty) = \infty$, the
classical trajectory of any massive particle is bounded, contrary to the
Schwarzchild solution

 For a massless particle, 
\be
({dr\over{d\lambda}})^2 = E^2 - \Lambda L^2 + {ML^2\over {r^2}}
\ee
The accessible region verifies ${1\over{r^2}} > {\Lambda L^2-E^2\over{ML^2}}$
and so, contrary to massive particles, some photons (not all) can go to
infinity. For a fixed energy, there is an upper bound $L_*$ such that when
$L>L_*$, the photon is confined to a bounded portion of the space .

\subsection{Existence of the planckian problem}

Q.F.T in a B.T.Z geometry and related thermodynamic properties have been extensively studied \cite{Das,Lee,Ca1,Ca2,Ca3,Ca4,Ca5,Ca6,Emp,Sen}.
In the Schwarzchild geometry, the Eddington-Finkelstein coordinates are the
best in which the planckian problem is revealed \cite{Br1}.As the essential of the Hawking radiation takes place in the
s channel,one usually study the slices $ \theta = C^{st},\phi = C^{st}$.
A photon which is detected at infinity after a time corresponding to the
life time of the black hole must have crossed the surface of the star with
a momentum which is so big that free scalar field description is dubious : this
is the transplanckian problem. Put in formula, a photon whose typical energy
is $0(M^{-1})$ reach a distance $x=0(M)$ from the horizon after a time
$v=0(M^3)$ only if it crossed the star surface with a momentum $ p_{st} \sim 0(M^{-1}e^{M^2})$
which is very large for a typical black hole.
\par What is the situation for the BTZ geometry? 
\par One rapidly finds \cite{BTZ2} 
\be
{^v_u\rbrace} = t \pm {1\over{2\sqrt{M}}} \ln |{r-\sqrt{M/\Lambda}\over
{r+\sqrt{M/\Lambda}}}|
\ee
so that the metric in the ingoing Eddington-Finkelstein coordinates becomes
\be
ds^2 = (\Lambda r^2-M)dv^2-2dv dr
\ee
The $s$ wave equation
\be
\partial_r [(\Lambda r^2-M)\partial_r + 2\partial_v] \phi(v,r) = 0
\ee
can be obtained from the hamiltonian
\be
H = (\Lambda r^2-M)p^2+2p_v p
\ee
        To derive the presence of the transplanckian problem for the B.T.Z geometry, we shall follow closely \cite{Br1}.The Hamiltonian equations of motion
are
\begin{equation}
  r^{\prime} = 2 ( \Lambda r^2 - M ) p + 2 p_v
\end{equation}

\begin{equation}
  p^{\prime} = - 2 \Lambda r p^2 \,\,\,\,\,\,\,  v^{\prime} = 2 p \,\,\,\,\,\,\, p_v = -\omega
\end{equation}
           Using on shell mass condition,one finds 
\[ r(v) = \sqrt{( \frac{2 \omega}{p(v)} + M) \frac{1}{\Lambda}} \]

\begin{eqnarray}
p(v) & = &  {\frac{p_{st}}{2\,{e^{v\,{\sqrt{M\,\Lambda }}}}}} - 
  {\frac{\omega }{M}} + {\frac{\omega }
    {2\,{e^{v\,{\sqrt{M\,\Lambda }}}}\,M}} + 
  {\frac{{\sqrt{{\frac{M\,{p_{st}^2} + 2\,p_{st}\,\omega }{M}}}}}
    {2\,{e^{v\,{\sqrt{M\,\Lambda }}}}}}      \nonumber\\
& + & {\frac{{e^{v\,{\sqrt{M\,\Lambda }}}}\,{{\omega }^2}}
   {2\,{M^2}\,\left( p_{st} + {\frac{\omega }{M}} + 
       {\sqrt{{\frac{M\,{p_{st}^2} + 2\,p_{st}\,\omega }{M}}}} \right) 
     }}               
\end{eqnarray}

     $ p_{st} $ is the value taken by $p$ when $ v = 0 $ ;it is the momemtum 
with which the photon crosses the horizon.

\par Let's give an estimate of the BTZ black hole life time in a very
na\"\i ve way. First of all, the total mass-energy of the black hole is not
 $M$. Actually, one can define masses in two different but
related ways. Like in \cite{BTZ2} , one may introduce the displacements at
infinity $N^{\infty},N^{\phi}$ and identify as mass $M$ and angular momentum
$J$ their conjugates variables in the hamiltonian formalism. But, when studying
stellar structures, it is natural to define as mass ``$m$'' the volume integral
of the total energy density. One then finds that the two quantities are related \cite{Lu}:
\be
m = M+1
\ee
   The quantity $ M $ is dimensionless; it  gives the mass using as unit    $ (G \sqrt{\Lambda}/c^2 )^{-1} $. 
\par As the typical energy of a photon is $0(T\sim\sqrt{M\Lambda})$, one finds
(neglecting back reaction) that the total number of photons the star can radiate
is $ 0(\frac{M+1}{\sqrt{ \Lambda M}})      $. As the typical time corresponding
to a photon is $0((\Lambda M)^{-1/2})$, the black hole will desappear after a
time
\be
T_{B.H} = 0(\frac{M+1}{\sqrt{ \Lambda M}}  ( M \Lambda )^{-1/2} ) = ({M+1\over{M\Lambda}})
\ee
so, for very massive BTZ black holes, the life-time depends only on the
cosmological constant.Detailed analysis should take into account the back reaction ; we nevertheless believe
formula (13) gives a correct order of magnitude.Two points should however be clarified.The first one is only a technical
question:what is the fraction of  energy  emmited in s- waves? In the Schwarzchild case ,it is $ 90 pc $.Altough we have not
completed the computation,equations (4) and (3) show that compared to the Scharzchild geometry,less modes can attain  spatial infinity.The same formula also reveals that for a given energy ,the s photons are priviledgied.So,one can reasonably
conjecture that the two dimensional (radial) model still exhibits the most important physical features.The second point is
a more fundamental one.To give an estimate of the typical energy of a photon in the asymptotic region,we used the equipartition theorem with the absolute temperature $ T=2 \sqrt{M\Lambda}$.One knows that in a curved background thermodynamic temperature $ T$ should be defined as a space-time dependent object\cite{Lan} ;then the  quantity $ T \sqrt{g_{00}}$ is the same in the hole geometry.In the asymptotic Scharzchild region,$ g_{00} = 1 $ and so all static observers in this zone measure the same temperature .For B.T.Z  geometry,$ g_{00}$ is an increasing function of the distance:the situation is different . We shall assume that $ T = 2 \sqrt{M \Lambda} $  is measured by an observer whose distance to the horizon is of the order of the black hole radius ;this is already the case for Scharzchild's solution. We do not impose  reflexive conditions on the boundary  and so we discard the cases where an equilibrium can be attained
\cite{Gi}.  
\par Now, consider a photon which crosses the star surface at $ v = 0 $ (with momentum $ p_{st} $ ). When it reaches a distance comparable to the black hole radius
$ (x = r - \sqrt{\frac{M}{\Lambda}}) $ after a time of the order of the black hole life time $ T_{B.H}= \frac{M + 1}{M \Lambda} $, its frequency typically
equals the black hole temperature $ \omega = 2 \sqrt{M \Lambda} $.Putting all this in
equation ($11.b$),one finds
\begin{eqnarray}
p_{st}  & = & {e^{{\frac{\left( 1 + M \right) \,{\sqrt{M\,\Lambda }}}
        {M\,\Lambda }}}}\,{\sqrt{M\,\Lambda }} - 
  {\frac{2\,{\sqrt{M\,\Lambda }}}{M}} + 
  {\frac{{\sqrt{M\,\Lambda }}}
    {{e^{{\frac{\left( 1 + M \right) \,{\sqrt{M\,\Lambda }}}
           {M\,\Lambda }}}}\,M}} + 
  {\frac{{e^{{\frac{\left( 1 + M \right) \,{\sqrt{M\,\Lambda }}}
           {M\,\Lambda }}}}\,{\sqrt{M\,\Lambda }}}{M}}  \nonumber\\
& + &   {\frac{{\sqrt{2 + M}}\,{\sqrt{\Lambda }}}
    {{e^{{\frac{\left( 1 + M \right) \,{\sqrt{M\,\Lambda }}}
          {M\,\Lambda }}}}}} - 
  {e^{{\frac{\left( 1 + M \right) \,{\sqrt{M\,\Lambda }}}
        {M\,\Lambda }}}}\,{\sqrt{2 + M}}\,{\sqrt{\Lambda }} + 
  {\frac{{\sqrt{M\,\Lambda }}}
    {{e^{{\frac{\left( 1 + M \right) \,{\sqrt{M\,\Lambda }}}
          {M\,\Lambda }}}}}}   
\end{eqnarray}

For a fixed cosmological constant, this goes to infinity as the mass
of the black hole increases. This is not the only planckian problem in BTZ.
\par One can give to $M$ a reasonable value and take a very small cosmological
constant : the momentum of a Hawking photon crossing the star surface is
again too large and our description is no more reliable.This kind of transplanckian problem is not present in the Schwarzchild geometry.Let us also note from equations(11) that the photon hugs the horizon.
It is straightforward that the truncation given by
\be
H = -g(p) [2\sqrt{M\Lambda} x g(p) + 2p_v]
\ee
with the choice $g(p) = \left\lbrace\begin{array}{ll}
p \ for\ p<1\\
1\ for\ p\geq 1
\end{array}
\right.$
avoids transplanckian problems like in \cite{Br1} : photons do not hug 
the horizon anymore.
\par This is a classical approach. We need a quantum confirmation in which
the classical trajectories  will be given by the path of the
saddle of the wave packets in the truncated theory.As explained in the introduction,we'll
not use  equation (15) but shall rather apply K.M.M theory.\\ 

\par\ To complete the analysis,we point out 
the dependence  of the effective potential on the angular momentum .
One important difference between BTZ and Schwarzchild geometries lies in
the fact that, as the former possess a non null Ricci scalar, minimal and
conformal gravitationnal couplings are different. However, $R$ being constant,
this gravitationnal coupling is strictly equivalent to a mass term and will be
discarded.The derivation of the effective potential is easy and directly reveals the field asymptotic behaviour  which can also be inferred from 
 the exact solution.
\par The separation of variables given by
\be
\phi = r^{1/2} e^{il\theta} e^{-i\omega t} R_{\omega l}(r)
\ee
reduces the Klein-Gordon equation in the BTZ configuration to
\be
{d^2R_{\omega l}(r)\over{dr^2_{\ast}}} + V(r) R_{\omega l} (r) = 0
\ee
with
\be
V(r) = \omega^2-\Lambda l^2+{M\Lambda\over 2} - {3\over 4} \Lambda^2r^2-{M^2
\over{r^2}} ({1\over 4}-l^2)
\ee
     This quantum potential is an unbounded increasing function of the distance .In the  Schwarzchild geometry  the torto\"\i se coordinate
\be
r_{\ast} = r+2M \ln({r\over{2M}}-1) 
\ee
verifies
\be
\lim_{r \rightarrow \infty} \frac{r_{\ast}}{ r} = 1
\ee
  In  the BTZ background ,one can invert the relation defining the torto\"\i se coordinate :
\be
r = \sqrt{M/\Lambda} \frac{1+ e^{2\sqrt{M\Lambda}r_*}}{1- e^{2\sqrt{M\Lambda}
r_*}}
\ee
so that at spatial infinity $r \sim  -{1\over{\Lambda r_{\ast}}}$ and 
$r_{\ast}  \sim 0^-$. The dominant term in the radial  potential reduces equation(17) to 
\be
{d^2 R_{\omega l}\over{dr^2_{\ast}}} - {3\over 4} {1\over {r^2_{\ast}}}
R_{\omega l} = 0
\ee
This equation is very different of the corresponding one in Schwarzchild
geometry
\be
{d^2R_{\omega l}\over{dr^2_{\ast}}} + \omega^2 R_{\omega l} = 0
\ee
\par The behaviour of the modes in the asymptotic BTZ region is
universal : it is, like in the Schwarzchild geometry, independent of the
characteristics $(M,\Lambda)$ of the black hole . In addition, it keeps no
trace of the frequency $\omega$ as shown in equation(22). The change of the
angular content of the theory when lowering the space-time dimension is encoded
in the potential through the term $l^2$ whose corresponding quantity is
$l(l+1)$ in a $3+1$ theory.
\par The equation (23), valid at spatial infinity, admitts simple solutions :
$r^{3/2}_{\ast}$ and $r^{-3/2}_{\ast}$. This gives two modes whose asymptotic
behaviours are opposite :
\par a) In the first case, $\phi \sim e^{il\theta} e^{-i\omega t} {1\over
{\Lambda^{3/2}r^2}}$ vanishes more rapidly than the Schwarzchild wave
function
\be
\phi \sim {1\over r} Y_{lm}(\theta,\phi) e^{i\omega (r+t)}
\ee
\par b) In the second case, $\phi \sim e^{il\theta} e^{-i\omega t} \Lambda^{1/2}$
doesn't decrease at spatial infinity .This was derived after an exact but lengthier calculation by \cite{Sen}.

\section{K.M.M theory in short}
      Kempf,Mangano and  Mann \cite{Kem1,Kem2,Kem3} replace the usual commutator relation by 
\be
[\hat{x},\hat{p}]= i \hbar (1 + \alpha \hat{x}^2 + \beta \hat{p}^2)
\ee
   This simple modification induces very profound consequences.The most important is the fact that the Heisenberg uncertainty  relation becomes

\be
  \Delta x \Delta p > \frac{\hbar}{2} (1 + \alpha <x>^2 + \alpha (\Delta x)^2 + \beta <p>^2 + \beta (\Delta p)^2)
\ee
          We restrict ourselves to $ \alpha = 0 $ .The motivation of this choice will be clear later.Solving the
Heisenberg inequality for $ \Delta p $,one easily obtains

\be
   (\Delta x )^2 \geq   \hbar^2  \beta + h^2 \beta^2 <p>^2
\ee
     The absolute minimum uncertainty in position $ \Delta x_0 = \hbar \sqrt{\beta} $ is attained when $<p> = 0$.So,the
new commutation relations imply that  position can not be known with arbitrary precision,{\em even if one increases the
uncertainty in momentum }.The case $ \alpha = 0 $ is interesting because it admitts a momentum representation.In fact,
on the domain $ S_{\infty}$ of functions decreasing faster than any power,the operator algebra is realised through
\begin{eqnarray}
   \hat{p}.\Psi(p) & = & p \Psi(p)  \\
   \hat{x}.\Psi(p) & = & i \hbar (1+ \beta p^2) \partial_p \Psi(p)
\end{eqnarray}

       $ S_{\infty} $ must be endowed with a Hilbert structure.The scalar product must be chosen so that $\hat{x}$ and
$\hat{p}$ are at least symmetric.This is the case if

\be
  <\Psi \vert \Phi> = \int_{-\infty}^{+ \infty} \frac{dp}{{1+ \beta p^2}} \Psi^{*}(p) \Phi(p)
\ee
         The normalised position eigenvectors,solutions of $ \hat{x} \Psi_\lambda = \lambda \Psi_\lambda $ are the wave
functions

\begin{equation}
  \Psi_\lambda (p) = \sqrt{\pi \sqrt{\beta}}  e^{- \frac{i \lambda}{\hbar \sqrt{\beta}} \arctan{\sqrt{\beta} p} }
\end{equation}
      The divergence of the quantity $ < \Psi_\lambda \vert \hat{p}^2 \Psi_\lambda > $ shows that these states have infinite energy 
and so are unphysical.Contrary to usual quantum theory,they cannot be attained as continous limits of states of finite energy
and decreasing uncertainty in position .By way of consequence ,there is no position representation of the operator algebra .
The scalar product of two distinct position eigenstates  
\begin{equation}
  < \Psi_\lambda \vert \Psi_\lambda^{\prime} > = \frac{\hbar \sqrt{\beta}}{ \pi (\lambda - \lambda^{\prime})} 
   \sin{\frac{ \pi (\lambda - \lambda^{\prime})} {2 \hbar \sqrt{\beta}}}
\end{equation}
is not necessarily null :the operators $ \hat{x} $ is not self adjoint but only symmetric; it admitts a  class of self adjoint extensions.
      To recover information on position , one defines the maximally localised states $ \vert \Psi_\xi^{ml} > $ as states in which
\begin{itemize}
   \item  the mean value of the position operator is $ \xi : < \Psi_\xi \vert \hat{x} \Psi_{\xi}>  = \xi $.
   \item  the uncertainty in position is minimal $ (\Delta x) =  (\Delta x_0) = \hbar \sqrt{\beta} $
\end{itemize}   
     In the momentum representation one finds those state wave functions to be 
\be
  \Psi_\xi^{ml} (p) = \sqrt{2 \sqrt{\beta}/\pi}(1 + \beta p^2)^{-1/2}  e^{-\frac{i \xi}{\hbar \beta} \arctan{(\sqrt{\beta} 
        p}) }
\ee
   They are well defined and of finite energy.
       The quasi position representation is defined by
\be
    \phi(\xi) = < \Psi_\xi^{ml} \vert \phi >
\ee
     and  is interpreted as  the probability  for the particle to be maximally localised  at $\xi$. For a "plane
wave", one finds the dispersion relation

\be
    \lambda(E) = \frac{2 \pi \hbar \sqrt{\beta}}{\arctan{ (\sqrt{2 m \beta E} ) }}
\ee 
   so that a minimal wave length  appears :$ \lambda_0 = 4 \hbar \sqrt{\beta} $.In this representation,the action of the operators is given by   
\begin{eqnarray}
   \hat{p} \Psi(\xi) & = &  \frac{ \tan{(- i \hbar \sqrt{\beta} \partial_\xi})}{\sqrt{\beta}}  \Psi(\xi)  \\
   \hat{x} \Psi(\xi) & = &  ( \xi + \sqrt{\beta} \tan{(- i \hbar \sqrt{\beta} \partial_\xi}))  \Psi(\xi)
\end{eqnarray} 

and the scalar product reads

\be
 < \Psi \vert \Phi> = \int_{- \infty}^{\infty} d\xi   \int_{- \infty}^{\infty} d\xi^{\prime}  \Psi^{\ast}(\xi)      \Psi(\xi^{\prime})e^{ i \arctan{(( \xi - \xi^{\prime})\hbar/\sqrt{\beta})}}   
\ee

\section{ Application to Black Hole physics}

     To begin this section let us emphasize that there should not be an oscillating term ,contrary to what was claimed in \cite{Br2}.For this let us review the main points of the analysis performed in that paper:

\begin{enumerate}
   \item  The wave function $ \Phi_\Omega (\theta) $ was computed in the  momentum representation.($ \Omega = 4 M \omega $              and $ \theta = \arctan{p}$)
   \item  The position wave function $ \chi_{\Omega} 
(x) = A_\Omega \int_0^{\pi/2} d\theta e^{i \theta x}                                           \Phi_\Omega(\theta)$ was worked out.
   \item  It was found that the right member of the position wave equation 
                 \be
                     (x \tan{(- i \partial_x )}- \Omega) \chi 
                 \ee
          was not zero but $ A_\Omega  e^{i \pi/2 }x$.A tentative was then made to explain this intriguing fact.
\end{enumerate}
     The most important characteristic of  K.M.M theory when $ \beta \not = 0 $ is the inexistence of a physically meaningfull position representation. As explained in the preceding section,we only have a quasi position representation.As can be inferred from (34),this is related to the  momentum one by

\be
     \chi_{\Omega} (x) = A_\Omega \int_0^{\pi/2} d\theta  \cos(\theta) e^{i \theta x}  \Phi_\Omega(\theta)
\ee
  
If the transform is well defined(and K.M.M show explicitely this to be the case), this is naturally a solution of the wave equation in the quasi position representation.{\em But,in this representation,the position operator is not simply the multiplication by x},( see formula (37)) and so the relevant equation is not (39).In brief,the oscillating term appeared
due to misleading in representations.Let us handle the problem more carefully.

\subsection{Schwarzchild geometry}  
       The  s wave equation of a complex scalar field in the Scharzchild background is
\be
  (i \hat{x} \hat{p} - 2 i \omega (\hat{x} + 2 M)) \chi_\omega (x) = 0
\ee 
  $\omega $  being the frequency.Its quantum counterpart  in the momentum representation reads
\be
  ( i \hbar (1 + \beta p^2) - 4 M \omega ) \chi_\omega (p) + i \hbar (1 + \beta p^2)(p - 2 \omega) \partial_p \chi_\omega (p) = 0
\ee

        Contrary to what is usually done in the litterature \cite{Br1},\cite{Br2} we did not restrict ourselves to the region near
the horizon:this results in an extra term  in  equation (41). One may solve this
equation  and use the Fourrier like transform specified by equation(34) to obtain the quasi position wave function, but the integral is not trivial.We can
directly consider the quasi position representation wave equation 

\be
  (( \xi \frac{\hbar}{\gamma} - 2 \omega \frac{\gamma}{\hbar}) \tan{(- i \gamma \partial_\xi )} + (\tan{(- i \gamma        \partial_\xi )})^2 - 2 \omega \xi - 4 M \omega ) \chi_{\omega} (\xi) = 0
\ee
  with $ \gamma = \hbar \sqrt{\beta} $ the minimal length uncertainty.
       This is an infinite order differential equation.
A priori it is difficult to handle. In particular,its order suggests
its general solution involves an infinity of arbitrary constants. Actually,there is a recipe which facilitates the analysis.
The minimal length $\gamma $ present in the tangent argument plays a crucial role when one expands the operators in their 
Taylor series. In fact,the differential operator acting  on $ \chi_\omega (\xi) $ is analytic in $\gamma$. If we could write
a similar expansion for the wave function,then we would obtain a differential equation for each order. This would mean that
we have replaced one infinite order equation by an illimited but numerable set of finite order ones. It is physically
resonable to expect that the K.M.M wave function dependance on the minimal length is a smooth one. This is the case for
the harmonic oscillator. One is then tempted to use
\be
  \chi_{\omega} (\xi) = f_0 (\xi) + \gamma f_1 (\xi) + \gamma^2  f_2 (\xi) + \cdots
\ee
        But,when studing trajectories,we are interested in the phase of the wave function.So,it is more interesting to
choose the parametrisation
\be
  \chi_{\omega} (\xi) = e^{f_0 (\xi) + \gamma f_1 (\xi) + \gamma^2  f_2 (\xi) + \cdots}
\ee
       The s wave equation then takes the form
\be
      \sum_{n=0}^{\infty}  \gamma^n  eq(n) = 0
\ee
        We give here the first members of this family of differential equations
\begin{equation}
eq(0) = -i\,\left( 4\,M\,\omega  + 2\,\xi \,\omega  \right)  + 
  \xi \,{f_0}'(\xi ) 
\end{equation}
\begin{equation}
 eq(1) =     \xi   f_{1}'(\xi)  
\end{equation}
\begin{eqnarray}
eq(2) = -2\,\omega \,{f_0}'(\xi ) + \xi \,{f_2}'(\xi ) + 
  i\,\left( -{{{f_0}'(\xi )}^2} - {f_0}''(\xi ) \right) 
		 \nonumber\\
 +   {\frac{\xi \,\left( -2\,{f_0}'(\xi )\,{f_0}''(\xi ) - 
       {f_0}'(\xi )\,\left( {{{f_0}'(\xi )}^2} + 
          {f_0}''(\xi ) \right)  - {f_0}^{(3)}(\xi )
        \right) }{3}} 
\end{eqnarray}
\begin{eqnarray}
eq(3)  & = &  -2\,\omega \,{f_1}'(\xi ) + \xi \,{f_3}'(\xi ) + 
  i\left( -2\,{f_0}'(\xi )\,{f_1}'(\xi ) - 
     {f_1}''(\xi ) \right)   \nonumber\\
& + &  {\frac{\xi \,\left( -\left( {f_1}'(\xi )\,
          \left( {{{f_0}'(\xi )}^2} + 
            {f_0}''(\xi ) \right)  \right)  - 
       {f_0}'(\xi )\,\left( 2\,{f_0}'(\xi )\,
           {f_1}'(\xi ) + {f_1}''(\xi ) \right) 
       \right) }{3}}   \nonumber\\
& + &  {\frac{\xi \,\left( -2\,
        \left( {f_1}'(\xi )\,{f_0}''(\xi ) + 
          {f_0}'(\xi )\,{f_1}''(\xi ) \right)  - 
       {f_1}^{(3)}(\xi ) \right) }{3}}
\end{eqnarray}
\begin{eqnarray}
eq(4) & = &  {\frac{2\,\omega \,{{{f_0}'(\xi )}^3}}{3}} + 
  {\frac{2\,i\,{{{f_0}'(\xi )}^4}}{3}} + 
  {\frac{2\,\xi \,{{{f_0}'(\xi )}^5}}{15}} - 
  i \,{{{f_1}'(\xi )}^2} - 
  \xi \,{f_0}'(\xi )\,{{{f_1}'(\xi )}^2} \nonumber\\
& - &  2\,\omega \,{f_2}'(\xi ) - 
  2\,i\,{f_0}'(\xi )\,{f_2}'(\xi ) - 
  \xi \,{{{f_0}'(\xi )}^2}\,{f_2}'(\xi ) + 
  \xi \,{f_4}'(\xi )   \nonumber\\
& + &  2\,\omega \,{f_0}'(\xi )\,{f_0}''(\xi ) + 
  4\,i\,{{{f_0}'(\xi )}^2}\,{f_0}''(\xi ) + 
  {\frac{4\,\xi \,{{{f_0}'(\xi )}^3}\,{f_0}''(\xi )}
    {3}} - \xi \,{f_2}'(\xi )\,{f_0}''(\xi )           \nonumber\\
& + &  2\,i\,{{{f_0}''(\xi )}^2} + 
  2\,\xi \,{f_0}'(\xi )\,{{{f_0}''(\xi )}^2} - 
  \xi \,{f_1}'(\xi )\,{f_1}''(\xi ) - 
  i\,{f_2}''(\xi )    -  \xi \,{f_0}'(\xi )\,{f_2}''(\xi ) \nonumber\\
& + & {\frac{2\,\omega \,{f_0}^{(3)}(\xi )}{3}} + 
  {\frac{8\,i\,{f_0}'(\xi )\,{f_0}^{(3)}(\xi )}{3}} + 
  {\frac{4\,\xi \,{{{f_0}'(\xi )}^2}\,
      {f_0}^{(3)}(\xi )}{3}}   +    {\frac{4\,\xi \,{f_0}''(\xi )\,{f_0}^{(3)}(\xi )}
    {3}}   \nonumber\\
& - & {\frac{\xi \,{f_2}^{(3)}(\xi )}{3}} + 
  {\frac{2\,i\,{f_0}^{(4)}(\xi )}{3}} + 
  {\frac{2\,\xi \,{f_0}'(\xi )\,{f_0}^{(4)}(\xi )}
    {3}} + {\frac{2\,\xi \,{f_0}^{(5)}(\xi )}{15}}
\end{eqnarray}

Two observations are important for the resolution:
\begin{itemize} 
   \item  the function $ f_n (\xi) $ appears for the first time in $ eq(n)$,through its first derivative
   \item  $eq(n)$ doesn't involve $ f_m (\xi) $ for $ m > n$ :
\end{itemize}
\be
    f_n^{\prime} (\xi) = F_n( f'_0 (\xi),f''_0 (\xi), ...,  f'_{n-1} (\xi),
f''_{n-1} (\xi) ,....        )
\ee

      Solving $ eq(0) $,we find $ f_0 (\xi)$.Putting this in $ eq(1)$ we compute $f_1 (\xi) $ ,etc.... Each contribution $ f_n (\xi) $ is determined up to a constant $ C_n$  but these constants do not  appear in the equations  of the $ f_m $ for $m > n $.Two different choices of this family of
constants give two  wave functions  which differ by a miltiplicative factor and so yields the same physics.For convenience
we put all the $C_i$ equal to zero.
        
One remarks that the odd contributions can consistently be put equal to zero : $ f_1 (\xi) = f_3 (\xi) = ... = 0 $.This is due to the fact that the equation (43)
is invariant under the reversal of the  $ \gamma $  sign.
         This analysis is close to perturbation theory.In fact,perturbation theory expresses the wave function of the modified(perturbed) system as an entire series in the coupling constant.This role is assumed in our situation by the minimal length 
coming from the new commutation relations.The analogy is pertinent since the zero order solution  is effectively
the solution of the "unperturbed" system defined by $ \gamma = 0$.As in pertrbation theory,the convergence of the serie is a
technically difficult question:we'll not adress it.
        For simplicity,let's first retain only the first contribution of K.M.M theory to the habitual wave function:it is proportional to $\gamma^2$.Separating the modulus and the phase,we can write the mode as
\be
   \phi_\omega (\xi,v) = e^{ a(\xi) \omega + b(\xi) \omega^2} e^{i( (- v + c(\xi)) \omega + d(\xi) \omega^2 + e(\xi)     \omega^3)}    
\ee
with
\begin{equation}
  a(\xi) = \frac{- 2 M \gamma^2}{\xi^2}
\end{equation}
\begin{equation}
 b(\xi ) = {\frac{-8\,{M^2}\,{{\gamma }^2}}{{{\xi }^2}}} - 
  {\frac{8\,M\,{{\gamma }^2}}{\xi }} 
\end{equation}

\begin{equation}
 c(\xi ) ={\frac{-4\,M\,{{\gamma }^2}}{3\,{{\xi }^2}}} + 
  2\,\xi  + 4\,M\,\log (\xi )
\end{equation}
\begin{equation} 
d(\xi ) ={\frac{8\,{M^2}\,{{\gamma }^2}}{{{\xi }^2}}} + 
  {\frac{8\,M\,{{\gamma }^2}}{\xi }} 
\end{equation}
\begin{equation} 
e(\xi ) = {\frac{32\,{M^3}\,{{\gamma }^2}}{3\,{{\xi }^2}}} + 
  {\frac{32\,{M^2}\,{{\gamma }^2}}{\xi }} - 
  {\frac{8\,{{\gamma }^2}\,\xi }{3}} - 
  16\,M\,{{\gamma }^2}\,\log (\xi )
\end{equation}

      Let us emphasize that $c(\xi)$ is the only one which does't vanish with $ \gamma $.The  new wave function differs significantly from the old one near the origin where it vanishes due to the $ e^{-1/ xi^2} $ factor.In the asymptotic region the inverse powers of $ \xi $ can be neglected for frequences near $ 1/M $ and so one recovers the habitual behaviour.
      Let us now study the classical trajectory associated with the gaussian wave paquet
\be
  \phi_(\xi,v) = \int_{-\infty}^{\infty} d\omega e^{- \frac{(\omega - \omega_0)^2}{\sigma^2}} \phi_\omega (\xi,v)
\ee
   Due to equation (53),the integrand is the product of two factors:the modulus and the phase.For a fixed point $ (\xi,v)$
of space time,two frequences are important:

\begin{itemize}
   \item  the frequency $ \omega_1 (\xi,v)$ for which the modulus is maximal
   \item  the frequency $ \omega_2 (\xi,v)$ for which the phase is stationnary
\end{itemize}
    If $\omega_1$ and $ \omega_2 $ are very different ,the gaussian integral is negligeable.In fact,the contributions of the neighberhood of $\omega_1 $ in the integral have important absolute values but the phase oscillates rapidly and those
contributions essentially cancel.The same conclusion holds for the neighberhood of $\omega_2$ :the phase varies slowly,the
contributions interfer in a constructive way.But their modulus are small and the result is not significant.The wave function
is significantly different from zero at the points $ (\xi,v) $  for which $ \omega_1 (\xi,v) \sim \omega_2 (\xi,v)$.
It can consequently be characterised  by the equality
$\omega_1 (\xi,v) = \omega_2 (\xi,v)$.The modulus of the integrand possesses a maximum only if
\be
    \sigma^2  b(\xi) < 1
\ee
    for all physical $ \xi $;it is located at

\be
   \omega_1 (\xi,v) = - \frac{ a(\xi) + \frac{2 \omega_ 0}{\sigma^2}}{2( - \frac{1}{\sigma^2} + b(\xi) )}
\ee 
       The phase is stationnary at
\be
    \omega_2 (\xi,v) = \frac{- d(\xi) \pm \sqrt{d(\xi)^2 - 3 (-v + c(\xi)) e(\xi)}}{3 e(\xi)}
\ee
        The preceding discussion then gives the trajectory without explicit computation of the integral:
\be
     v(\xi) = c(\xi) - \frac{d(\xi)^2}{3 e(\xi)}+
 \frac{1}{3 e(\xi)} \left(d(\xi)- \frac{ 3 e(\xi) ( a(\xi)+ \frac{2 \omega_0}{\sigma^2}) }{  2(- \frac{1}{\sigma^2} + 
  b(\xi))}\, \right)^2
\ee
     A trivial analysis shows that when $\gamma $ goes to zero,the first term is the only one which survives .One then
simply recovers the well known trajectory of habitual Minkowski wave paquets.
     We saw that the condition $ \sigma^2 b(\xi) < 1 $ is necessary for the real factor of the integrand to be bounded.As this must be fulfilled for all positive $\xi$,we conclude that the notion of trajectory doesn't apply to gaussian paquets 
with characteristics outside a given region in the plane $(\omega_0,\sigma)$.The non locality of such wave paquets 
manifests itself in the entire  space time.
    From equation (62)  it is obvious that the frequency $\omega_2$  is real provided
\be
   v >  sign(e(\xi)) \left( 3 c(\xi) - \frac{ d(\xi)^2}{e(\xi)} \right)
\ee
    When this inequality is not satisfied,the phase is nowhere stationnary on the real axis of the $\omega$.As we argued,this would imply destructive interference almost everywhere on this axis.The wave function is then  deprived of   a peak  and the notion of trajectory is lost in this region.We also saw that the condition $ \sigma^2 b(\xi) < 1 $ is necessary for the modulus to be bounded.Calling $ \xi_1 > \xi_2 $ the zeros of  $ \sigma^2 b(\xi) = 1 $ ,one finds:
\begin{enumerate}
   \item  If $ \sigma < \frac{1}{\sqrt{ \gamma}} $, inequality (61) is verified
           by all real $\xi$.The region of non locality  is composed of the
           space time points $(\xi,v)$ which violate the inequality $(64)$.
   \item  If  $ \sigma \geq \frac{1}{\sqrt{ \gamma}} $,the non locality region
           satisfies the supplementary condition $ \xi < \xi_1 $. A trivial       computation shows that $ \xi_1 < 0 $. By way of consequence,in this case non locality manifests itself inside the horizon.Then the falling of the partner to the emitted photon on the singularity would have no signification.This may in
principle put a new light on the unitarity problem,{\em without advocating the
complementarity hypothesis} \cite{Eng},\cite{Sus}.
\end{enumerate}

For a null value of $ \gamma$ ,$e(\xi)$ vanishes and the last  equation  shows that the zone of non locality is empty.

      What is the effect of higher order corrections?The forth order contribution to the phase is
\begin{eqnarray}
 \varphi_4 & = &   {\frac{32\,{{\gamma }^4}\,\xi \,{{\omega }^5}}{5}} + 
  {\frac{-32\,M\,{{\gamma }^4}\,{{\omega }^4} - 
      256\,{M^2}\,{{\gamma }^4}\,{{\omega }^5}}{\xi }}  \nonumber\\
& + &   {\frac{{\frac{-8\,M\,{{\gamma }^4}\,{{\omega }^3}}
       {3}} - 96\,{M^2}\,{{\gamma }^4}\,{{\omega }^4} - 
     256\,{M^3}\,{{\gamma }^4}\,{{\omega }^5}}{{{\xi }^
      2}}}             \nonumber\\
& + &  {\frac{8\,M\,{{\gamma }^4}\,{{\omega }^2} - 
     {\frac{32\,{M^2}\,{{\gamma }^4}\,{{\omega }^3}}
       {3}} - 128\,{M^3}\,{{\gamma }^4}\,
      {{\omega }^4} - {\frac{512\,{M^4}\,{{\gamma }^4}\,
         {{\omega }^5}}{3}}}{{{\xi }^3}}}    \nonumber\\
& + &  {\frac{{\frac{53\,M\,{{\gamma }^4}\,\omega }{15}} + 
     20\,{M^2}\,{{\gamma }^4}\,{{\omega }^2} - 
     {\frac{32\,{M^3}\,{{\gamma }^4}\,{{\omega }^3}}
       {3}} - 64\,{M^4}\,{{\gamma }^4}\,{{\omega }^4} - 
     {\frac{256\,{M^5}\,{{\gamma }^4}\,{{\omega }^5}}
       {5}}}{{{\xi }^4}}}      \nonumber\\
& + &  64\,M\,{{\gamma }^4}\,{{\omega }^5}\,\log (\xi )
\end{eqnarray}
while the corresponding quantity for the modulus is
\begin{eqnarray}
m_4  & = &  {\frac{32\,M\,{{\gamma }^4}\,{{\omega }^4}}{\xi }} + 
 {\frac{16\,M\,{{\gamma }^4}\,{{\omega }^3} + 
      96\,{M^2}\,{{\gamma }^4}\,{{\omega }^4}}{{{\xi }^
       2}}}                 \nonumber\\
& + &  {\frac{8\,M\,{{\gamma }^4}\,{{\omega }^2} + 
     64\,{M^2}\,{{\gamma }^4}\,{{\omega }^3} + 
     128\,{M^3}\,{{\gamma }^4}\,{{\omega }^4}}{{{\xi }^
      3}}}                \nonumber\\
& - &  {\frac{2\,M\,{{\gamma }^4}\,\omega  + 
     20\,{M^2}\,{{\gamma }^4}\,{{\omega }^2} + 
     64\,{M^3}\,{{\gamma }^4}\,{{\omega }^3} + 
     64\,{M^4}\,{{\gamma }^4}\,{{\omega }^4}}{{{\xi }^
      4}}}
\end{eqnarray}
        The preceding  analysis is accurate when these contributions can be neglected when compared to the second order's ones. As shown in the figures,this
is not true  near the horizon where the wave function tends to the constant zero. So ,altough our method for identifying the  region where non localisability sets in  is  qualitatively correct, our approximation  of the solution makes 
quantitavly irrelevant  the analysis when one is too close to the  horizon.In the 
asymptotic region, the correction to the phase and the modulus  becomes negligeable  for typical Hawking photons: one then verify the remarquable
conjecture \cite{Br2} that roughly speaking, the modification of the commutation relations acts essentially near the horizon.

       Let us remark that singular solutions in $ \gamma $ may appear when dealing with an equation implying a series . For simplicity, if one considers
the equation $ (\gamma^2 \partial_\xi^2 + \gamma \partial_\xi + m^2) \phi = 0 $,
then a solution of the form $ e^{ \gamma^(-1) f_{-1}(\xi) + f_0 + \gamma f_{1}(\xi) + ...} $ exhists,with $ f_{-1}(\xi)$ linear.As we already argued,such
solutions should be discarded .
\section{B.T.Z geometry}
     The s mode wave obeys the equation
\begin{equation}
    ( \Lambda \hat{x}^2 \hat{p} + 2 \sqrt{ M \Lambda} \hat{x} \hat{p} - 2 \omega) \, | \, \chi_{\omega}> = 0  
\end{equation}
    In the  quasi position representation,this becomes 
\begin{eqnarray}
  (  \Lambda  (\xi + \gamma \tan{(- i \gamma \partial_\xi)})^2 \gamma^{-1} \tan{(- i \gamma \partial_\xi)}  +   \nonumber\\
 2 \sqrt{M \Lambda} (\xi + \gamma \tan{(- i \gamma \partial_\xi )}) \gamma^{-1} \tan{(- i \gamma \partial_\xi)} - 2 \omega) \chi_{\omega} (\xi) = 0
\end{eqnarray}
The analysis proceeds exactly like in the previous section.The expressions are simply lengthier,in part  due to the supplementary parameter $\Lambda$.We expand the wave function as in equation (45) but we now take null odd contributions from the begining .This gives the equations $ eq(0), eq(2),eq(4) $  and their solutions listed in the appendix. 
\section{Conclusion}
     The equation we quantised doesn't treat on the same footing the variables $x,v$ and so breaks explicitely the covariance. This was also
the case in \cite{Un1},\cite{Br1},\cite{Br2} where it was proved to be unavoidable for solving the transplanckian problem. Since K.M.M
itself is a frame dependant formulation,the picture seems better.Second,the upsetting of the quantum mechanics rules may
change the mathematical description of particles.This would result in a modification of the Hawking emission characteristics and should be cleared up. Although we have worked basically with the first non vanishing correction,we conjecture this describes the essential physics. In fact $ \gamma $ must be very tiny to avoid significant deviations from usual quantum mechanics for well known cases like the oscillator,or the  Hydrogen  atom.When analysing a gaussian wave pacquet at this order,the non locality zone exhibited a dependance on its mean frequency and  dispersion.Altough  higher order corrections may quantitatively change this  near the horizon, our computations strongly suggest that the  non locality zone may cover a part of the black hole interior.If this is true, then the partner of the Hawking photon does't merge with the singularity. This may be promising for the unitarity problem but a more reliable treatment should include back reaction and
use a better approximation of the wave function near the horizon. 
    Last but not least, if there is a minimal length,a star should not 
collapse to a point:it would stop at least when it's radius  reaches this length.
In this case the Hawking radiation would rapidly cease.But,as the gravitationnal  collapse is a purely classical mechanism,the details of such a halting mechanism
is not trivial.
\bigskip

\underline{Acknowledgement} It is a pleasure to thank Philippe Spindel,R.Brout , F.Englert and C.Gabriel for enlightening discussions.

\section{ Appendix}
\begin{equation}
eq(0) = -2\,\omega  - i\,\left( 2\,{\sqrt{M\,\Lambda }}\,\xi  + 
     \Lambda \,{{\xi }^2} \right) \,{f_0}'(\xi )
\end{equation}

\begin{eqnarray}
      eq(2) & = &   - \,\Lambda \,{f_0}'(\xi ) - 
  2\,\,{\sqrt{M\,\Lambda }}\,{{{f_0}'(\xi )}^2} -
  \,\Lambda \,\xi \,{{{f_0}'(\xi )}^2} + 
  {\frac{2\, i \,{\sqrt{M\,\Lambda }}\,\xi \,
      {{{f_0}'(\xi )}^3}}{3}}    \nonumber\\
& + &  {\frac{\left( i \,\Lambda \,{{\xi }^2}\,
        {{{f_0}'(\xi )}^3} \right) }{3}} - 
  2\,i\,{\sqrt{M\,\Lambda }}\,\xi \,({f_2})'(\xi ) - 
  i\,\Lambda \,{{\xi }^2}\,{f_2}'(\xi ) - 
  2\,\,{\sqrt{M\,\Lambda }}\,{f_0}''(\xi )   \nonumber\\
& - &  2\,\,\Lambda \,\xi \,({f_0})''(\xi ) + 
  2\, i \,{\sqrt{M\,\Lambda }}\,\xi \,{f_0}'(\xi )\,
   {f_0}''(\xi ) + i \,\Lambda \,{{\xi }^2}\,
   {f_0}'(\xi )\,{f_0}''(\xi )    \nonumber\\
& + &  {\frac{2\, i\,{\sqrt{M\,\Lambda }}\,\xi \,
      {f_0}^{(3)}(\xi )}{3}} + 
  {\frac{i\,\Lambda \,{{\xi }^2}\,{f_0}^{(3)}(\xi )}{3}}
\end{eqnarray}
\begin{eqnarray}
  eq(4)&=&  \left( {\frac{4}{3}} + i \right) \,\Lambda \,
   {{({f_0})'(\xi )}^3} + 
  {\frac{4\,{\sqrt{M\,\Lambda }}\,{{{f_0}'(\xi )}^4}}{3}} + 
  {\frac{2\,\Lambda \,\xi \,{{{f_0}'(\xi )}^4}}{3}} - 
  {\frac{4\,i}{15}}\,{\sqrt{M\,\Lambda }}\,\xi \,
   {{{f_0}'(\xi )}^5}                  \nonumber\\
& + & {\frac{-2\,i}{15}}\,\Lambda \,{{\xi }^2}\,
   {{{f_0}'(\xi )}^5} - \Lambda \,{f_2}'(\xi ) - 
  4\,{\sqrt{M\,\Lambda }}\,{f_0}'(\xi )\,{f_2}'(\xi ) - 
  2\,\Lambda \,\xi \,{f_0}'(\xi )\,{f_2}'(\xi )  \nonumber\\
& + & 2\,i\,{\sqrt{M\,\Lambda }}\,\xi \,{{{f_0}'(\xi )}^2}\,
   {f_2}'(\xi ) + i\,\Lambda \,{{\xi }^2}\,
   {{({f_0})'(\xi )}^2}\,{f_2}'(\xi ) - 
  2\,i\,{\sqrt{M\,\Lambda }}\,\xi \,{f_4}'(\xi )   \nonumber\\
& - &  i\,\Lambda \,{{\xi }^2}\,{f_4}'(\xi ) + 
  \left( 4 + 3\,i \right) \,\Lambda \,{f_0}'(\xi )\,
   {f_0}''(\xi ) + 8\,{\sqrt{M\,\Lambda }}\,
   {{{f_0}'(\xi )}^2}\,{f_0}''(\xi ) \nonumber\\
& + &  6\,\Lambda \,\xi \,{{{f_0}'(\xi )}^2}\,{f_0}''(\xi ) - 
  {\frac{8\,i}{3}}\,{\sqrt{M\,\Lambda }}\,\xi \,
   {{({f_0})'(\xi )}^3}\,{f_0}''(\xi ) - 
  {\frac{4\,i}{3}}\,\Lambda \,{{\xi }^2}\,
   {{{f_0}'(\xi )}^3}\,{f_0}''(\xi )     \nonumber\\
& + & 2\,i\,{\sqrt{M\,\Lambda }}\,\xi \,{f_2}'(\xi )\,
   {f_0}''(\xi ) + i\,\Lambda \,{{\xi }^2}\,{f_2}'(\xi )\,
   {f_0}''(\xi ) + 4\,{\sqrt{M\,\Lambda }}\,
   {{{f_0}''(\xi )}^2}            \nonumber\\
& + &  4\,\Lambda \,\xi \,{{{f_0}''(\xi )}^2} - 
  4\,i\,{\sqrt{M\,\Lambda }}\,\xi \,{f_0}'(\xi )\,
   {{{f_0}''(\xi )}^2} - 
  2\,i\,\Lambda \,{{\xi }^2}\,{f_0}'(\xi )\,
   {{{f_0}''(\xi )}^2}                \nonumber\\
& - &  2\,{\sqrt{M\,\Lambda }}\,{f_2}''(\xi ) - 
  2\,\Lambda \,\xi \,{f_2}''(\xi ) + 
  2\,i\,{\sqrt{M\,\Lambda }}\,\xi \,{f_0}'(\xi )\,
   {f_2}''(\xi )         \nonumber\\
& + &  i\,\Lambda \,{{\xi }^2}\,{f_0}'(\xi )\,{f_2}''(\xi ) + 
  \left( {\frac{4}{3}} + i \right) \,\Lambda \,
   {f_0}^{(3)}(\xi ) + {\frac{16\,{\sqrt{M\,\Lambda }}\,
      {f_0}'(\xi )\,{f_0}^{(3)}(\xi )}{3}}    \nonumber\\
& + &  {\frac{14\,\Lambda \,\xi \,{f_0}'(\xi )\,{f_0}^{(3)}(\xi )}
    {3}} - {\frac{8\,i}{3}}\,{\sqrt{M\,\Lambda }}\,\xi \,
   {{{f_0}'(\xi )}^2}\,{f_0}^{(3)}(\xi )   \nonumber\\
& + &  {\frac{-4\,i}{3}}\,\Lambda \,{{\xi }^2}\,{{{f_0}'(\xi )}^2}\,
   {f_0}^{(3)}(\xi ) - {\frac{8\,i}{3}}\,
   {\sqrt{M\,\Lambda }}\,\xi \,{f_0}''(\xi )\,
 {f_0}^{(3)}(\xi )          \nonumber\\
& + &  {\frac{-4\,i}{3}}\,\Lambda \,{{\xi }^2}\,{f_0}''(\xi )\,
   {f_0}^{(3)}(\xi ) + {\frac{2\,i}{3}}\,
   {\sqrt{M\,\Lambda }}\,\xi \,{f_2}^{(3)}(\xi ) + 
  {\frac{i}{3}}\,\Lambda \,{{\xi }^2}\,{f_2}^{(3)}(\xi )       \nonumber\\
& + &  {\frac{4\,{\sqrt{M\,\Lambda }}\,{f_0}^{(4)}(\xi )}{3}} + 
  {\frac{4\,\Lambda \,\xi \,{f_0}^{(4)}(\xi )}{3}} - 
  {\frac{4\,i}{3}}\,{\sqrt{M\,\Lambda }}\,\xi \,
   {f_0}'(\xi )\,{f_0}^{(4)}(\xi )     \nonumber\\
& + &  {\frac{-2\,i}{3}}\,\Lambda \,{{\xi }^2}\,{f_0}'(\xi )\,
   {f_0}^{(4)}(\xi ) - {\frac{4\,i}{15}}\,
   {\sqrt{M\,\Lambda }}\,\xi \,{f_0}^{(5)}(\xi ) - 
  {\frac{2\,i}{15}}\,\Lambda \,{{\xi }^2}\,{f_0}^{(5)}(\xi )           
\end{eqnarray}

Retaining first the second order contribution, one defines functions $ a(\xi),b(\xi),...$ as in equations (54-58) and finds:  
\begin{equation}
      a(\xi)  = {\frac{-\left( {{\gamma }^2}\,{\sqrt{M\,\Lambda }} \right) }
    {2\,M\,\Lambda \,{{\xi }^2}}} + 
  {\frac{{{\gamma }^2}\,{\sqrt{M\,\Lambda }}}
    {2\,M\,\Lambda \,{{\left( {\frac{2\,{\sqrt{M\,\Lambda }}}
             {\Lambda }} + \xi  \right) }^2}}}
\end{equation}
\begin{equation}
      b(\xi)  = {\frac{-{{\gamma }^2}}{2\,M\,\Lambda \,{{\xi }^2}}} + 
  {\frac{{{\gamma }^2}\,{\sqrt{M\,\Lambda }}}
    {2\,{M^2}\,\Lambda \,\xi }} - 
  {\frac{{{\gamma }^2}}
    {2\,M\,\Lambda \,{{\left( {\frac{2\,{\sqrt{M\,\Lambda }}}
             {\Lambda }} + \xi  \right) }^2}}} - 
  {\frac{{{\gamma }^2}\,{\sqrt{M\,\Lambda }}}
    {2\,{M^2}\,\Lambda \,\left( {\frac{2\,{\sqrt{M\,\Lambda }}}
          {\Lambda }} + \xi  \right) }}
\end{equation}
\begin{equation}
       c(\xi)  = {\frac{-\left( {{\gamma }^2}\,{\sqrt{M\,\Lambda }} \right) }
    {3\,M\,\Lambda \,{{\xi }^2}}} + 
  {\frac{{{\gamma }^2}\,{\sqrt{M\,\Lambda }}}
    {3\,M\,\Lambda \,{{\left( {\frac{2\,{\sqrt{M\,\Lambda }}}
             {\Lambda }} + \xi  \right) }^2}}} + 
  {\frac{{\sqrt{M\,\Lambda }}\,\log (\xi )}{M\,\Lambda }} - 
  {\frac{{\sqrt{M\,\Lambda }}\,
      \log (2\,{\sqrt{M\,\Lambda }} + \Lambda \,\xi )}{M\,
      \Lambda }}
\end{equation}
\begin{equation}
       d(\xi)  = {\frac{{{\gamma }^2}}{2\,M\,\Lambda \,{{\xi }^2}}} - 
  {\frac{{{\gamma }^2}\,{\sqrt{M\,\Lambda }}}
    {{M^2}\,\Lambda \,\xi }} - 
  {\frac{{{\gamma }^2}\,{\sqrt{M\,\Lambda }}}
    {2\,{M^2}\,\Lambda \,\left( {\frac{2\,{\sqrt{M\,\Lambda }}}
          {\Lambda }} + \xi  \right) }} - 
  {\frac{3\,{{\gamma }^2}\,\log (\xi )}{4\,{M^2}}} + 
  {\frac{3\,{{\gamma }^2}\,
      \log ({\frac{2\,{\sqrt{M\,\Lambda }}}{\Lambda }} + \xi )}
      {4\,{M^2}}}
\end{equation}
\begin{eqnarray}
     e(\xi)  & = &   {\frac{{{\gamma }^2}\,{\sqrt{M\,\Lambda }}}
    {6\,{M^2}\,{{\Lambda }^2}\,{{\xi }^2}}} - 
  {\frac{{{\gamma }^2}}{2\,{M^2}\,\Lambda \,\xi }} - 
  {\frac{{{\gamma }^2}\,{\sqrt{M\,\Lambda }}}
    {6\,{M^2}\,{{\Lambda }^2}\,
      {{\left( {\frac{2\,{\sqrt{M\,\Lambda }}}{\Lambda }} + 
           \xi  \right) }^2}}} - 
  {\frac{{{\gamma }^2}}
    {2\,{M^2}\,\Lambda \,\left( {\frac{2\,{\sqrt{M\,\Lambda }}}
          {\Lambda }} + \xi  \right) }}               \nonumber\\
& + &  {\frac{-\left( {{\gamma }^2}\,{\sqrt{M\,\Lambda }}\,
        \log (\xi ) \right) }{2\,{M^3}\,\Lambda }} + 
  {\frac{{{\gamma }^2}\,{\sqrt{M\,\Lambda }}\,
      \log ({\frac{2\,{\sqrt{M\,\Lambda }}}{\Lambda }} + \xi )}
      {2\,{M^3}\,\Lambda }}
\end{eqnarray}

               The second non null contribution to the phase is
\begin{eqnarray}
      \varphi_4   & = &
      {\frac{53\,{{\gamma }^4}\,{\sqrt{M\,\Lambda }}\,\omega }
    {60\,M\,\Lambda \,{{\xi }^4}}} - 
  {\frac{53\,{{\gamma }^4}\,{\sqrt{M\,\Lambda }}\,\omega }
    {60\,M\,\Lambda \,{{\left( {\frac{2\,{\sqrt{M\,\Lambda }}}
             {\Lambda }} + \xi  \right) }^4}}} + 
  {\frac{5\,{{\gamma }^4}\,{{\omega }^2}}
    {4\,M\,\Lambda \,{{\xi }^4}}} - 
  {\frac{5\,{{\gamma }^4}\,{\sqrt{M\,\Lambda }}\,{{\omega }^2}}
    {6\,{M^2}\,\Lambda \,{{\xi }^3}}}                  \nonumber\\
& + &  {\frac{3\,{{\gamma }^4}\,{{\omega }^2}}{8\,{M^2}\,{{\xi }^2}}} - 
  {\frac{{{\gamma }^4}\,{\sqrt{M\,\Lambda }}\,{{\omega }^2}}
    {8\,{M^3}\,\xi }} + {\frac{{{\gamma }^4}\,{{\omega }^2}}
    {4\,M\,\Lambda \,{{\left( {\frac{2\,{\sqrt{M\,\Lambda }}}
             {\Lambda }} + \xi  \right) }^4}}} - 
  {\frac{{{\gamma }^4}\,{\sqrt{M\,\Lambda }}\,{{\omega }^2}}
    {6\,{M^2}\,\Lambda \,{{\left( {\frac{2\,{\sqrt{M\,\Lambda }}}
             {\Lambda }} + \xi  \right) }^3}}}        \nonumber\\
& + &  {\frac{-\left( {{\gamma }^4}\,{{\omega }^2} \right) }
    {8\,{M^2}\,{{\left( {\frac{2\,{\sqrt{M\,\Lambda }}}{\Lambda }} + 
           \xi  \right) }^2}}} + 
  {\frac{{{\gamma }^4}\,{\sqrt{M\,\Lambda }}\,{{\omega }^2}}
    {8\,{M^3}\,\left( {\frac{2\,{\sqrt{M\,\Lambda }}}{\Lambda }} + 
        \xi  \right) }} - {\frac{{{\gamma }^4}\,
      {\sqrt{M\,\Lambda }}\,{{\omega }^3}}{6\,{M^2}\,{{\Lambda }^2}\,
      {{\xi }^4}}} + {\frac{5\,{{\gamma }^4}\,{{\omega }^3}}
    {6\,{M^2}\,\Lambda \,{{\xi }^3}}}           \nonumber\\
& + &  {\frac{-11\,{{\gamma }^4}\,{\sqrt{M\,\Lambda }}\,{{\omega }^3}}
    {6\,{M^3}\,\Lambda \,{{\xi }^2}}} + 
  {\frac{15\,{{\gamma }^4}\,{{\omega }^3}}{4\,{M^3}\,\xi }} - 
  {\frac{{{\gamma }^4}\,{\sqrt{M\,\Lambda }}\,{{\omega }^3}}
    {3\,{M^2}\,{{\Lambda }^2}\,
      {{\left( {\frac{2\,{\sqrt{M\,\Lambda }}}{\Lambda }} + \xi 
            \right) }^4}}} - 
  {\frac{{{\gamma }^4}\,{{\omega }^3}}
    {6\,{M^2}\,\Lambda \,{{\left( {\frac{2\,{\sqrt{M\,\Lambda }}}
             {\Lambda }} + \xi  \right) }^3}}}               \nonumber\\
& + &  {\frac{7\,{{\gamma }^4}\,{\sqrt{M\,\Lambda }}\,{{\omega }^3}}
    {12\,{M^3}\,\Lambda \,{{\left( {\frac{2\,{\sqrt{M\,\Lambda }}}
             {\Lambda }} + \xi  \right) }^2}}} + 
  {\frac{5\,{{\gamma }^4}\,{{\omega }^3}}
    {2\,{M^3}\,\left( {\frac{2\,{\sqrt{M\,\Lambda }}}{\Lambda }} + 
        \xi  \right) }} - {\frac{{{\gamma }^4}\,{{\omega }^4}}
    {4\,{M^2}\,{{\Lambda }^2}\,{{\xi }^4}}} + 
{\frac{2\,{{\gamma }^4}\,{\sqrt{M\,\Lambda }}\,{{\omega }^4}}
    {3\,{M^3}\,{{\Lambda }^2}\,{{\xi }^3}}}     \nonumber\\
& + &  {\frac{-5\,{{\gamma }^4}\,{{\omega }^4}}
    {4\,{M^3}\,\Lambda \,{{\xi }^2}}} + 
  {\frac{5\,{{\gamma }^4}\,{\sqrt{M\,\Lambda }}\,{{\omega }^4}}
    {2\,{M^4}\,\Lambda \,\xi }} + 
  {\frac{{{\gamma }^4}\,{\sqrt{M\,\Lambda }}\,{{\omega }^4}}
    {6\,{M^3}\,{{\Lambda }^2}\,
      {{\left( {\frac{2\,{\sqrt{M\,\Lambda }}}{\Lambda }} + \xi 
            \right) }^3}}} + 
  {\frac{5\,{{\gamma }^4}\,{{\omega }^4}}
    {8\,{M^3}\,\Lambda \,{{\left( {\frac{2\,{\sqrt{M\,\Lambda }}}
             {\Lambda }} + \xi  \right) }^2}}}                   \nonumber\\
& + &  {\frac{15\,{{\gamma }^4}\,{\sqrt{M\,\Lambda }}\,{{\omega }^4}}
    {8\,{M^4}\,\Lambda \,\left( {\frac{2\,{\sqrt{M\,\Lambda }}}
          {\Lambda }} + \xi  \right) }} - 
  {\frac{{{\gamma }^4}\,{\sqrt{M\,\Lambda }}\,{{\omega }^5}}
    {20\,{M^3}\,{{\Lambda }^3}\,{{\xi }^4}}} + 
  {\frac{{{\gamma }^4}\,{{\omega }^5}}
    {6\,{M^3}\,{{\Lambda }^2}\,{{\xi }^3}}} - 
  {\frac{3\,{{\gamma }^4}\,{\sqrt{M\,\Lambda }}\,{{\omega }^5}}
    {8\,{M^4}\,{{\Lambda }^2}\,{{\xi }^2}}}        \nonumber\\
& + &  {\frac{7\,{{\gamma }^4}\,{{\omega }^5}}{8\,{M^4}\,\Lambda \,\xi }} + 
  {\frac{{{\gamma }^4}\,{\sqrt{M\,\Lambda }}\,{{\omega }^5}}
    {20\,{M^3}\,{{\Lambda }^3}\,
      {{\left( {\frac{2\,{\sqrt{M\,\Lambda }}}{\Lambda }} + \xi 
            \right) }^4}}} + 
  {\frac{{{\gamma }^4}\,{{\omega }^5}}
    {6\,{M^3}\,{{\Lambda }^2}\,
      {{\left( {\frac{2\,{\sqrt{M\,\Lambda }}}{\Lambda }} + \xi 
            \right) }^3}}}                    \nonumber\\
& + &  {\frac{3\,{{\gamma }^4}\,{\sqrt{M\,\Lambda }}\,{{\omega }^5}}
    {8\,{M^4}\,{{\Lambda }^2}\,
      {{\left( {\frac{2\,{\sqrt{M\,\Lambda }}}{\Lambda }} + \xi 
            \right) }^2}}} + 
  {\frac{7\,{{\gamma }^4}\,{{\omega }^5}}
    {8\,{M^4}\,\Lambda \,\left( {\frac{2\,{\sqrt{M\,\Lambda }}}
          {\Lambda }} + \xi  \right) }} + 
  {\frac{25\,{{\gamma }^4}\,{\sqrt{M\,\Lambda }}\,{{\omega }^3}\,
      \log (\xi )}{8\,{M^4}}}                     \nonumber\\
& + &   {\frac{35\,{{\gamma }^4}\,{{\omega }^4}\,\log (\xi )}{16\,{M^4}}} + 
  {\frac{7\,{{\gamma }^4}\,{\sqrt{M\,\Lambda }}\,{{\omega }^5}\,
      \log (\xi )}{8\,{M^5}\,\Lambda }} - 
  {\frac{25\,{{\gamma }^4}\,{\sqrt{M\,\Lambda }}\,{{\omega }^3}\,
      \log ({\frac{2\,{\sqrt{M\,\Lambda }}}{\Lambda }} + \xi )}{8\,
      {M^4}}}                 \nonumber\\
& + &  {\frac{-35\,{{\gamma }^4}\,{{\omega }^4}\,
      \log ({\frac{2\,{\sqrt{M\,\Lambda }}}{\Lambda }} + \xi )}{16\,
      {M^4}}} - {\frac{7\,{{\gamma }^4}\,{\sqrt{M\,\Lambda }}\,
      {{\omega }^5}\,\log ({\frac{2\,{\sqrt{M\,\Lambda }}}
          {\Lambda }} + \xi )}{8\,{M^5}\,\Lambda }}     
\end{eqnarray}
    while the corresponding quantity for the modulus is
\begin{eqnarray}
     m_4  & = &
     {\frac{-\left( {{\gamma }^4}\,{\sqrt{M\,\Lambda }}\,\omega  \right) }
    {2\,M\,\Lambda \,{{\xi }^4}}} + 
  {\frac{{{\gamma }^4}\,{\sqrt{M\,\Lambda }}\,\omega }
    {2\,M\,\Lambda \,{{\left( {\frac{2\,{\sqrt{M\,\Lambda }}}
             {\Lambda }} + \xi  \right) }^4}}} + 
  {\frac{5\,{{\gamma }^4}\,{{\omega }^2}}
    {4\,M\,\Lambda \,{{\xi }^4}}}        \nonumber\\
& + &   {\frac{-7\,{{\gamma }^4}\,{\sqrt{M\,\Lambda }}\,{{\omega }^2}}
    {6\,{M^2}\,\Lambda \,{{\xi }^3}}} + 
  {\frac{7\,{{\gamma }^4}\,{{\omega }^2}}{8\,{M^2}\,{{\xi }^2}}} - 
  {\frac{3\,{{\gamma }^4}\,{\sqrt{M\,\Lambda }}\,{{\omega }^2}}
    {4\,{M^3}\,\xi }} - {\frac{{{\gamma }^4}\,{\sqrt{M\,\Lambda }}\,
      {{\omega }^2}}{6\,{M^2}\,\Lambda \,
      {{\left( {\frac{2\,{\sqrt{M\,\Lambda }}}{\Lambda }} + \xi 
            \right) }^3}}}                    \nonumber\\
& + &  {\frac{{{\gamma }^4}\,{\sqrt{M\,\Lambda }}\,{{\omega }^2}}
    {8\,{M^3}\,\left( {\frac{2\,{\sqrt{M\,\Lambda }}}{\Lambda }} + 
        \xi  \right) }} + {\frac{{{\gamma }^4}\,
      {\sqrt{M\,\Lambda }}\,{{\omega }^3}}{{M^2}\,{{\Lambda }^2}\,
      {{\xi }^4}}} - {\frac{3\,{{\gamma }^4}\,{{\omega }^3}}
    {2\,{M^2}\,\Lambda \,{{\xi }^3}}} + 
  {\frac{3\,{{\gamma }^4}\,{\sqrt{M\,\Lambda }}\,{{\omega }^3}}
    {2\,{M^3}\,\Lambda \,{{\xi }^2}}} - 
  {\frac{5\,{{\gamma }^4}\,{{\omega }^3}}{4\,{M^3}\,\xi }}    
\nonumber\\
& + &   {\frac{{{\gamma }^4}\,{{\omega }^3}}
    {2\,{M^2}\,\Lambda \,{{\left( {\frac{2\,{\sqrt{M\,\Lambda }}}
             {\Lambda }} + \xi  \right) }^3}}} + 
  {\frac{{{\gamma }^4}\,{\sqrt{M\,\Lambda }}\,{{\omega }^3}}
    {{M^3}\,\Lambda \,{{\left( {\frac{2\,{\sqrt{M\,\Lambda }}}
             {\Lambda }} + \xi  \right) }^2}}} + 
  {\frac{5\,{{\gamma }^4}\,{{\omega }^3}}
    {4\,{M^3}\,\left( {\frac{2\,{\sqrt{M\,\Lambda }}}{\Lambda }} + 
        \xi  \right) }} + {\frac{{{\gamma }^4}\,{{\omega }^4}}
    {4\,{M^2}\,{{\Lambda }^2}\,{{\xi }^4}}}     \nonumber\\
& + &  {\frac{-\left( {{\gamma }^4}\,{\sqrt{M\,\Lambda }}\,
        {{\omega }^4} \right) }{2\,{M^3}\,{{\Lambda }^2}\,{{\xi }^3}}
    } + {\frac{5\,{{\gamma }^4}\,{{\omega }^4}}
    {8\,{M^3}\,\Lambda \,{{\xi }^2}}} - 
  {\frac{5\,{{\gamma }^4}\,{\sqrt{M\,\Lambda }}\,{{\omega }^4}}
    {8\,{M^4}\,\Lambda \,\xi }} + 
  {\frac{{{\gamma }^4}\,{{\omega }^4}}
{4\,{M^2}\,{{\Lambda }^2}\,
      {{\left( {\frac{2\,{\sqrt{M\,\Lambda }}}{\Lambda }} + \xi 
            \right) }^4}}}             \nonumber\\
& + & {\frac{{{\gamma }^4}\,{\sqrt{M\,\Lambda }}\,{{\omega }^4}}
    {2\,{M^3}\,{{\Lambda }^2}\,
      {{\left( {\frac{2\,{\sqrt{M\,\Lambda }}}{\Lambda }} + \xi 
            \right) }^3}}} + 
  {\frac{5\,{{\gamma }^4}\,{{\omega }^4}}
    {8\,{M^3}\,\Lambda \,{{\left( {\frac{2\,{\sqrt{M\,\Lambda }}}
             {\Lambda }} + \xi  \right) }^2}}} + 
  {\frac{5\,{{\gamma }^4}\,{\sqrt{M\,\Lambda }}\,{{\omega }^4}}
    {8\,{M^4}\,\Lambda \,\left( {\frac{2\,{\sqrt{M\,\Lambda }}}
          {\Lambda }} + \xi  \right) }}           \nonumber\\
& + & {\frac{-5\,{{\gamma }^4}\,\Lambda \,{{\omega }^2}\,\log (\xi )}
    {16\,{M^3}}} + {\frac{5\,{{\gamma }^4}\,\Lambda \,{{\omega }^2}\,
      \log ({\frac{2\,{\sqrt{M\,\Lambda }}}{\Lambda }} + \xi )}{16\,
      {M^3}}}
\end{eqnarray}
          Once again , the second non vanishing contribution  can not de neglected near the origin.

\section{Figures  Captions}
\bigskip

\noindent Fig 1 \,\,\,\,\,\, Plot of the typical momentum $ p_{st} $ of a photon crossing the star surface of a B.T.Z black hole in terms of the cosmological constant $ \Lambda $ for a small mass $ M =  1 $.
\bigskip

\noindent Fig 2 \,\,\,\,\,\,  Plot ,in the Schwarchild geometry, of the trajectories of a wave paquet ( of mean frequency $ \omega_0 = 1/M $ ,dispersion $ \sigma = \omega_0/10 $  and mass  $ M = 10^{40} $ ) for two values of the minimal
  length $ \gamma = 1,10 $ .The trajectories  are  stopped at the begining of the zone of  non locality. 

\bigskip

\noindent Fig 3 \,\,\,\,\,\, Plot of  the  same trajectories in the asymptotic region.The dependence on the minimal length has desappeared.
\bigskip

\noindent Fig 4 \,\,\,\,\,\ Plot of the quotient of the  forth contribution to  the second one for the wave function  phase. The  analysis based solely on the 
second order terms is seen to be true far from the horizon.

\noindent Fig 5 \,\,\,\,\,\,  Plot ,in the  B.T.Z  geometry, of the trajectories of a wave paquet ( of mean frequency $ \omega_0 = 1/M $ ,dispersion $ \sigma = \omega_0/10 $  and mass  $ M = 1 $ ) for two values of the minimal
  length $ \gamma = 1,5 $.The trajectories  are  stopped at the begining of the zone of  non locality. 
\bigskip

\noindent Fig 6 \,\,\,\,\,\, Plot of  the  same trajectories in the asymptotic region.The dependence on the minimal length has desappeared.
\bigskip

\noindent Fig 7 \,\,\,\,\,\ Plot of the quotient of the  forth contribution to  the second one for the wave function  phase.

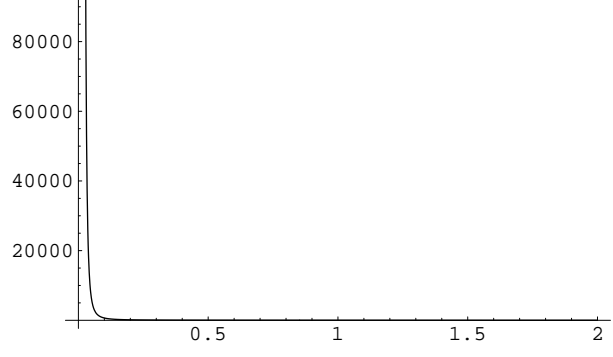
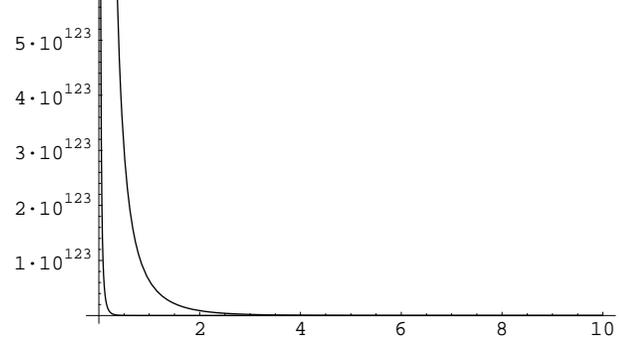
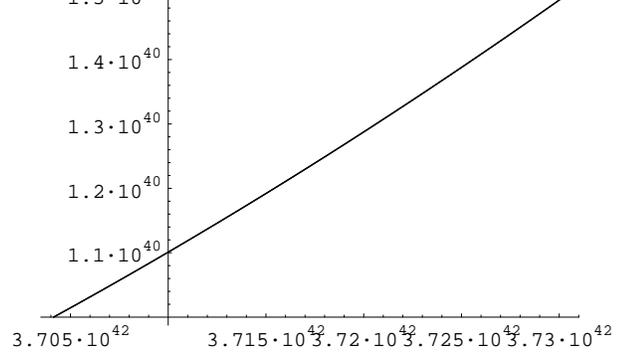
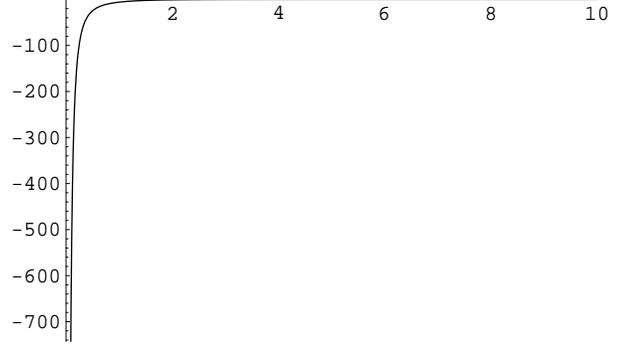
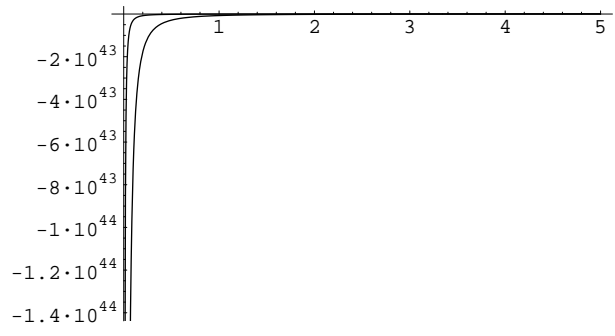
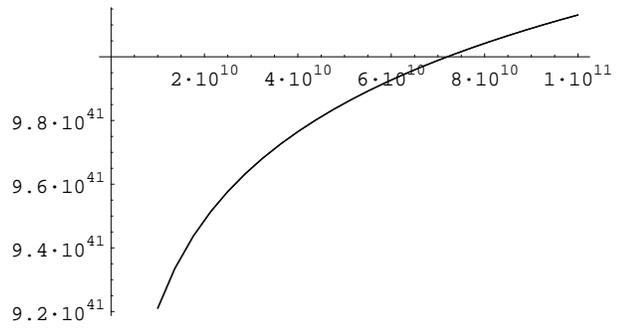
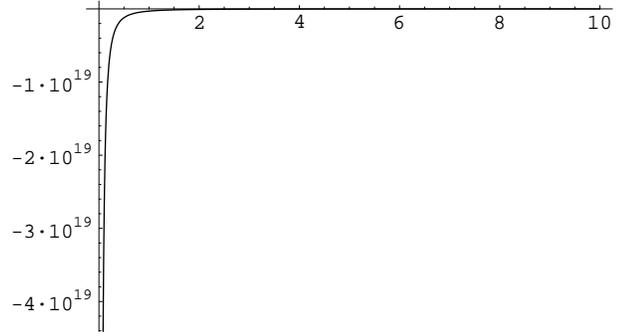
\end{document}